\documentclass[aps,pra,reprint,superscriptaddress,floatfix]{revtex4-2}

\usepackage{graphicx}
\usepackage{dcolumn}
\usepackage{bm}

\usepackage{amsmath}
\usepackage{amssymb}
\usepackage{lipsum}
\usepackage{diagbox}
\usepackage{physics}
\usepackage{bbding}
\usepackage{hyperref}
\usepackage{orcidlink}

\newcommand{\creation}{\hat{a}^\dagger}
\newcommand{\anihilation}{\hat{a}}
\newcommand{\hamc}{\hat{\mathcal{H}}}
\newcommand{\ham}{\hat{H}}
\newcommand{\bcreation}{\hat{b}^\dagger}
\newcommand{\banihilation}{\hat{b}}

\begin{document}


\title{Dipolar optimal control of entangled current states}

\author{Héctor Briongos-Merino \orcidlink{0009-0003-7859-6270}}
\affiliation{Departament de F\'isica Qu\`antica i Astrof\'isica,
Facultat de F\'{\i}sica, Universitat de Barcelona, 08028
Barcelona, Spain.}
\affiliation{Institut de Ci\`encies del Cosmos de la Universitat de
Barcelona, ICCUB, 08028 Barcelona, Spain.}
\author{Felipe Isaule \orcidlink{0000-0003-1810-0707}}
\affiliation{Instituto de Física, Pontificia Universidad Católica de Chile, Avenida Vicuña Mackenna 4860, Santiago, Chile.}
\author{Bruno Juli\'a-D\'iaz \orcidlink{0000-0002-0145-6734}}
\affiliation{Departament de F\'isica Qu\`antica i Astrof\'isica,
Facultat de F\'{\i}sica, Universitat de Barcelona, 08028
Barcelona, Spain.}
\affiliation{Institut de Ci\`encies del Cosmos de la Universitat de
Barcelona, ICCUB, 08028 Barcelona, Spain.}
\author{Montserrat Guilleumas \orcidlink{0000-0003-2011-1480}}
\affiliation{Departament de F\'isica Qu\`antica i Astrof\'isica, Facultat
de F\'{\i}sica, Universitat de Barcelona, 08028 Barcelona, Spain.}
\affiliation{Institut de Ci\`encies del Cosmos de la Universitat de
Barcelona, ICCUB, 08028 Barcelona, Spain.}

\date{\today}

\begin{abstract}
Quantum state control is a fundamental tool for quantum technologies. In this work, we propose and analyze the use of quantum optimal control to exploit the dipolar interaction of ultracold atoms on a lattice ring, focusing on the generation of selected states with entangled circulation. This scheme requires time-dependent control over the orientation of the magnetic field, a technique that is feasible in ultracold atom laboratories. The system's evolution is driven by just two independent control functions. We describe the symmetry constraints of this approach and numerically test them using the extended Bose-Hubbard model. We find that the proposed control can engineer entangled current states with perfect fidelity across a wide range of systems, and that in the remaining cases, the theoretical upper bounds for fidelity are reached.
\end{abstract}

\maketitle


\section{\label{sec:Introduction}Introduction}

Efficient control of quantum systems represents one of the fundamental prerequisites for multiple quantum technologies, such as quantum computing \cite{farhi_quantum_2001, chen_gapped_2009}, simulation \cite{sorensen_adiabatic_2010, ljubotina_optimal_2022}, and metrology \cite{lin_optimal_2021}. However, such control presents a challenge because of the intrinsic fragility of quantum states. To tackle this problem, numerous approaches for the engineering of quantum states have been proposed, spanning from integrable systems \cite{grun_protocol_2022} and adiabatic techniques \cite{jarzynski_generating_2013, venuti_relaxation_2017, hatomura_shortcuts_2024} to more sophisticated methods based on the principles of Quantum Optimal Control (QOC) theory \cite{albertini_notions_2002, dong_quantum_2010, glaser_training_2015, dalessandro_introduction_2021, koch_quantum_2022}.

QOC provides a general methodology to engineer control protocols. It typically involves manipulating external fields to drive a quantum system toward a desired target state or operation. These protocols are often designed to optimize specific performance metrics, such as maximizing fidelity while minimizing the energy consumption, or maximizing robustness against noise and system errors.
The controlled external fields have been employed to tune the interparticle interaction \cite{morandi_optimal_2024}, to displace \cite{wu_optimal_2025, morandi_optimal_2024} or modulate \cite{haug_machinelearning_2021, perciavalle_quantum_2024} the trapping potential.
Several algorithms fall under the QOC framework \cite{machnes_comparing_2011, koch_quantum_2022}, including gradient-ascent pulse engineering (GRAPE) \cite{khaneja_optimal_2005}, chopped random basis (CRAB) optimization \cite{muller_one_2022}, Krotov's method \cite{reich_monotonically_2012}, and machine learning-based techniques \cite{carleo_machine_2019, haug_machinelearning_2021, norambuena_physicsinformed_2024}.

Among quantum systems, atomtronic circuits based on ultracold atoms appear as an ideal platform for testing QOC. They offer a high degree of tunability and control, as well as potential for developing novel quantum devices \cite{amico_roadmap_2021, amico_colloquium_2022}. Within these, ultracold dipolar atoms~\cite{lu_strongly_2011, tang_bose_2015,aikawa_boseeinstein_2012,griesmaier_boseeinstein_2005,miyazawa_boseeinstein_2022} and molecules~\cite{bohn_cold_2017, langen_quantum_2024, bigagli_observation_2024} are a specially interesting platform in quantum technologies. They interact through dipole-dipole interactions, which can be precisely manipulated using external electromagnetic fields. This makes them excellent candidates for quantum control.
In this direction, it has been demonstrated that dipolar systems give rise to rich physics in atomtronic circuits~\cite{gallemi_role_2013, rovirola_ultracold_2024, ymai_tunneling_2025}. In particular, persistent circulation can be induced by appropriately tuning the orientation of the dipoles \cite{briongos-merino_dipolar_2025}. This phenomenon relies on the magnetostirring technique, which has been developed to experimentally generate angular momentum and, consequently, vortices in dipolar quantum gases \cite{klaus_observation_2022,prasad_vortex_2019, prasad_arbitraryangle_2021, bland_vortices_2023}.

In this work, we combine the QOC tools with the dipolar magnetostirring method to propose and validate a new driving mechanism. We consider a small number of polar bosons confined in an atomtronic lattice ring [Fig.~\ref{fig:system}(a)], a natural architecture for generating persistent currents, and drive the state into a target entangled current (EC) state using magnetostirring. The magnetostirring protocol is updated to perform state preparation by simply optimizing the orientation of an external electromagnetic field over time [Fig.~\ref{fig:system}(b)]. We have chosen to prepare EC states as they represent a generalization of the current states that are relevant for atomtronic applications as rotation sensors \cite{gustavson_precision_1997, pelegri_quantum_2018} and qubit proposals \cite{mooij_phaseslip_2005}.
Later, we derive the set of reachable EC states under this driving method, describing the boundaries that prevent the preparation of some of the EC states in  systems with an even number of sites. Furthermore, we compute
optimal trajectories for the preparation of entangled currents in the lattice ring, demonstrating that the resulting fidelities are in agreement with the theoretical predictions.

This work is organized as follows. Section \ref{sec:QOC} introduces the system, the control scheme and the target states. Section \ref{sec:limitations} analyzes the capabilities of the proposed protocol from a controlability point of view, and defines symmetry-based boundaries to its performance. Section \ref{sec:Dipolarcontrol} illustrates the feasibility of the protocol by numerically obtaining optimal stirring trajectories for symmetry-constrained and unconstrained systems. We discuss the applicability in the regime of {hard-core} bosons in Sec. \ref{sec:hardcore} and comment on the experimental feasibility in Sec \ref{sec:experimental}. Finally, Sec. \ref{sec:Conclusion} presents the conclusions.

\section{\label{sec:QOC}Dipolar quantum optimal control}
 Given a quantum system under control, the Hamiltonian takes the form
\begin{equation}
\label{eq:Hamiltonian}
    \hamc (t) = \ham_0 + \ham_c(\boldsymbol{\theta}(t))\,,
\end{equation}
where $\ham_0$ is the drift part and $\ham_c(\boldsymbol{\theta}(t))$ represents the control Hamiltonian, which depends on a set of time-dependent parameters $\boldsymbol{\theta}(t)$. QOC aims to modulate these parameters over a time $t_c$ to drive the system toward a desired target operation.

\begin{figure}[tb]
    \centering
    \includegraphics[width=0.9\linewidth]{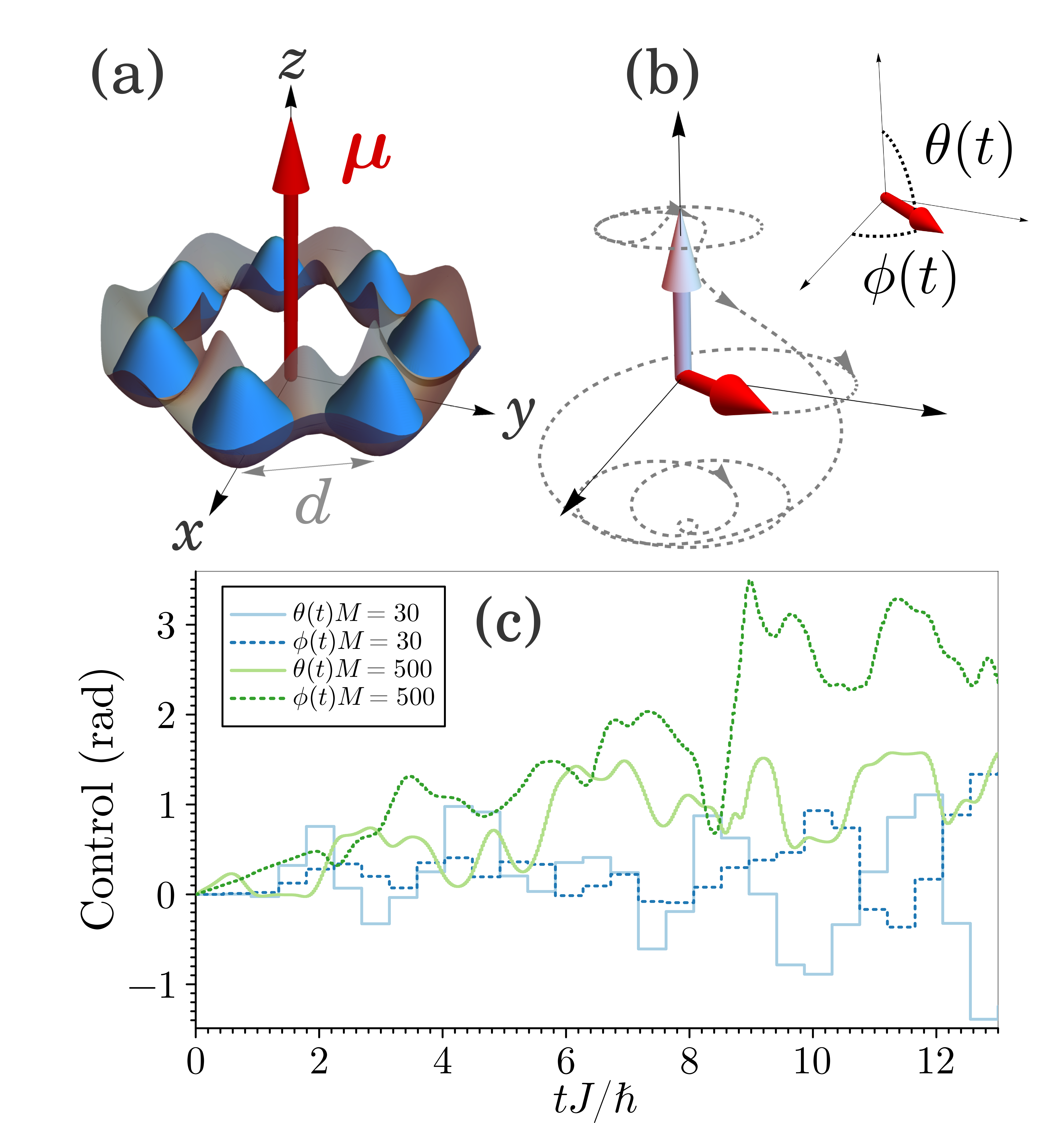}
    \caption{(a) Schematic representation of the system with $L=7$ sites and dipole polarization along the $z$-axis. (b) Example trajectory during a dipolar QOC protocol. (c) Example of optimal trajectories for two different numbers of steps $M$, preparing the $\Omega=\{-1,0,1\}$ EC state for $N=2$, $L=5$.}
    \label{fig:system}
\end{figure}

Our model system is a lattice ring circuit composed of $L$ sites and loaded with $N$ dipolar bosonic atoms. 
Within the tight-binding approximation, the system is described by the extended dipolar Bose-Hubbard Hamiltonian (dBHH). 
Therefore, the drift Hamiltonian includes the kinetic and on-site interaction terms
\begin{equation}
    \ham_0 =  \sum_{j=1}^{L}\left[-J(\hat a^\dagger_{j+1} \hat a_{j}+\hat a^\dagger_{j} \hat a_{j+1}) + \frac{U}{2} \hat{n}_j(\hat{n}_j-1) \right] \,,
\end{equation}
where $\hat{a}_j (\hat{a}_j^\dag)$ are the bosonic annihilation (creation) operators for the $j$-th site, $\hat{n}_j=a^\dagger_{j}\hat a_{j}$ is the particle number operator of site $j$, $J$ is the nearest neighbors tunneling strength and $U$ is the on-site interaction strength. As depicted in Fig.~\ref{fig:system}(a), we consider that the ring circuit lies on the plane $z=0$, with equally spaced wells by a lattice spacing $d$ and at the same distance from the origin. In turn, our work takes as control parameter the orientation of the global dipole moment, 
taking advantage of the anisotropic nature of the dipole-dipole interaction to drive the system. 
Thus, the control part of the Hamiltonian is given by the long-range, anisotropic dipolar interaction. It reads, 
\begin{equation}
\label{eq:dipolarcontrol}
    \ham_c(\boldsymbol{\mu}, t) = U_d \sum_{\substack{j=1\\k>j}}^{L} \left[\frac{1}{|\mathbf{r}_{jk}|^3}-3\frac{\left(\boldsymbol{\mu}(t) \cdot \mathbf{r}_{jk}\right)^2}{|\boldsymbol{\mu}|^2|\mathbf{r}_{jk}|^5}\right]\,\hat{n}_j\hat{n}_k\,,
\end{equation}
where $U_d$ is the dipolar interaction strength, $\boldsymbol{\mu}(t)$ is the atomic magnetic dipole moment and $\mathbf{r}_{jk}=\mathbf{r}_j-\mathbf{r}_k$ is the vector that points from the $k$-th to the $j$-th site. The total number of degrees of freedom of the control is two, as the orientation of $\boldsymbol{\mu}(t)$ is completely determined by the two spherical angles [see Fig.~\ref{fig:system}(b)]. Then, for the rest of this work, the set of time-dependent parameters is $\boldsymbol{\theta}(t) = \{\theta(t), \phi(t)\}$. These controls are discretized in steps as illustrated in Fig. \ref{fig:system}(c). Thus, using a high number of steps $M$ results in a continuous control (green lines), while a small number of steps results in a  discrete one (blue lines). 

In this work, we aim to drive a quantum system from an initial state $\ket{\Psi_0}$ to a target state $\ket{\Psi_{\mathrm{T}}}$ by dynamically changing the orientation of the magnetic dipole moment. This is typically achieved by maximizing the state fidelity
\begin{equation}
    F[\boldsymbol{\theta}(t)] = \left| \bra{\Psi_{\mathrm{T}}} U(\boldsymbol{\theta}(t); t_c)\ket{\Psi_0} \right|^2, 
\end{equation}
where $\ket{\Psi(t_c)} = U(\boldsymbol{\theta}(t); t_c)\ket{\Psi_0}$ is the state of the system at final time $t_c$, and $U(\boldsymbol{\theta}(t); t_c)$ is the unitary evolution operator generated by the time-dependent Hamiltonian $\hat{\mathcal{H}}(\boldsymbol{\theta}(t))$. The state preparation problem can thus be formulated as the maximization of the fidelity functional $F[\boldsymbol{\theta}(t)]$ over the set of time-dependent trajectories $\boldsymbol{\theta}(t)$.

For all the state preparation protocols considered in this work, we choose the initial state $\ket{\Psi_0}$ to be the ground state of the system when the dipole moment is aligned along the $z$-axis $\boldsymbol{\mu}=\mu \, \mathbf{e}_z$.
In a coherent atomtronic ring, current states
are quantized by a winding number $k$ associated with the quasimomentum. 
Therefore, a single-particle current state
reads
\begin{equation}
\label{eq:Winding}
    \ket{k} = \hat{b}_k^\dagger \ket{\operatorname{vac}} = \frac{1}{\sqrt{L}}\sum_{j=1}^{L} e^{i 2\pi k j / L} \hat{a}_j^\dagger \ket{\operatorname{vac}} \, ,
\end{equation}
where $\hat{b}_k^\dagger$ is the creation operator for the quasimomentum $k$. 
Thus, following Ref.~\cite{haug_machinelearning_2021}, we choose entangled currents (EC) states as the target states of the QOC protocol. The target EC states correspond to superpositions of winding modes of the form
\begin{equation}
    \ket{\Psi_{\mathrm{EC}}} = \frac{1}{\sqrt{K N !}}\sum_{k\in\Omega} \left(\hat{b}_k^\dagger\right)^{N} \ket{\operatorname{vac}} \,,
\end{equation}
where $\Omega=\{k_1, k_2, \dots, k_{K}\}$ is a set of $K$ winding numbers that appear in the entangled state. 
In the following, we focus on two representative cases:
the NOON state ($K = 2$) and the W state ($K = 3$). We include the preparation of a different family of entangled states in Appendix \ref{ap:spatiallyentangled}. 

\section{\label{sec:limitations}Controlability of the system}

A fundamental question in QOC is whether, or to what extent, it is possible to control a quantum system to achieve any physically permitted evolution. This is called complete controllability \cite{ramakrishna_controllability_1995, schirmer_complete_2001}, and refers to systems with control Hamiltonians such that any unitary operator $U$ can be achieved through time evolution under a Hamiltonian of the form~(\ref{eq:Hamiltonian}). 
This question about controllability has been investigated by previous works \cite{jurdjevic_control_1972, ramakrishna_controllability_1995, turinici_quantum_2001, chakrabarti_quantum_2007} on linearly controlled systems of the form $\hat{H}_c(\boldsymbol{\theta}(t))=\sum_m \theta_m(t) \hat{H}_m $,
where $\theta_m(t)$ are independent control functions.
Such works have found that finite-dimensional systems are completely controlable if and only if the Lie subalgebra generated by the set $\{\hat{H}_0,\dots,\hat{H}_m\}$ coincides with the one associated with the unitary group $U(N_\mathcal{H})$, where $N_\mathcal{H}=\operatorname{dim} \mathcal{H}$ is the dimension of the Hilbert space \cite{ramakrishna_controllability_1995, schirmer_complete_2001, fu_complete_2001, zeier_symmetry_2011, dalessandro_introduction_2021}. Thus, computing the dimension of the Lie subalgebra is sufficient to determine whether the system is completely controllable. Such computations can be performed numerically \cite{schirmer_complete_2001, zeier_symmetry_2011}.

There is no simple algebraic condition to verify controllability for the nonlinear control Hamiltonian considered in this work.
Nevertheless, we can assess the system's controllability in two stages. First, we use the results developed for systems with independent linear controls to set an upper bound on the controllability. Second, we employ numerical simulations to demonstrate that our dipolar control scheme saturates this bound, and therefore conclude that the controllability aligns with the linear control predictions. This second step presents limitations due to the intractable size of possible target states, and therefore we will restrict the specific validation to the EC states presented before. 

For the first stage, we have computed the dimension of the Lie algebra generated by the drift Hamiltonian and a set of control operators $\{\hat{c}_j\}$ that can be identified from the control Hamiltonian (\ref{eq:dipolarcontrol}) to define an upper bound to the controllability of our model. 
To construct this set, we have considered all pairs of number operators ($\hat{n}_j\hat{n}_k$) and examined whether there exists any other pair that has the same weight for all values of the control $\boldsymbol{\mu}$. For a ring with an odd number of sites, each pair is independent, resulting in a set $\{\hat{c}_j\}$ with $L (L-1) / 2 $ operators of the form $\hat{n}_j\hat{n}_k$, with $j\neq k$. For an even number of sites, the system exhibits a symmetry under spatial inversion, $\mathbf{r}\to\mathbf{-r}$, which reduces the number of independent operator pairs. Specifically, the pairs $\hat{n}_j\hat{n}_k$ and $\hat{n}_{j+L/2}\hat{n}_{k+L/2}$ have the same weight for all values of $\boldsymbol{\mu}$, and cannot be considered independent control operators. In this case, the set of control operators $\{\hat{c}_j\}$ have terms of the form $\hat{n}_j\hat{n}_k + \hat{n}_{j+L/2}\hat{n}_{k+L/2}$. 

Table \ref{tab:liealgebra} presents the results obtained for the dimension of the Lie algebra for different system sizes. It shows that systems with an even number of sites cannot be completely controllable under our dipolar control, as the dimension of the generated Lie algebra is significantly smaller than the full unitary group $U(N_{\mathcal{H}})$ dimension. In contrast, systems with an odd number of sites generate all the elements of the unitary group and do not present any controllability limitations in the linear independent case. This discards fundamental controlability limits coming from control operator limitations.
The limitation in the even case arises because the evolution preserves the system's inversion symmetry, thus not all unitary transformations can be applied with dipolar QOC in such systems.

\begin{table}[tbp]
    \centering
    \begin{tabular}{c|c c c c} \hline \hline
    \backslashbox{$N$}{$L$} & 3 & 4       & 5        & 6       \\ \hline
        2 & 36/36                 & 33/100  & 225/225  & 208/441   \\
        3 & 100/100               & 199/400 & 1225/1225 & X       \\
        4 & 225/225               & 617/1225& X       & X          \\
        5 & 441/441               & X     &    X     &  X       \\
        6 & 784/784               &  X    &     X    &   X       \\ 
        7 & 1296/1296              &  X    &      X   &    X     \\ \hline \hline
    \end{tabular}
    \caption{Numerically obtained Lie algebra dimension out of the full $U(N_\mathcal{H})$ algebra dimension for a system with $L$ sites and $N$ bosons, assuming independent controls. As the algorithm's computational cost scales with $(\operatorname{dim}\mathcal{H})^8$, only the smallest systems can be analyzed.}
    \label{tab:liealgebra}
\end{table}

Moreover, in systems invariant under inversion symmetry ($L$ even), the Lie algebra of the system is reduced to a subalgebra of the unitary group, preventing full unitary controllability \cite{zeier_symmetry_2011}. In the context of state transfer, this symmetry divides the Hilbert space into two disjoint subspaces, corresponding to even (eigenvalue +1) and odd (eigenvalue -1) parity states under the inversion transformation. When applying the inversion symmetry to the labeling of the sites ($j\to j+L/2$), the single-particle winding number states [Eq.~(\ref{eq:Winding})] transform under the inversion operator $\hat{I}$ as $\hat I \ket{k} = e^{-i\pi k}  \ket{k}$. The transformation of the target EC states is given by:
\begin{equation}
    \hat{I}\ket{\Psi_{\mathrm{EC}}} = \frac{1}{\sqrt{K N!}}\sum_{k\in\Omega} e^{-i\pi N k} \left(\hat{b}_k^\dagger\right)^{N} \ket{\operatorname{vac}} \, .
\end{equation}
From this expression, we observe that any EC state built with an even number of particles exhibits even parity under the inversion transformation. Even-parity states can also exist for odd particle numbers, provided that the set $\Omega$ contains only even values the winding number $k$. Conversely, states with odd parity under $\hat{I}$ correspond to configurations with an odd number of particles and a set $\Omega$ containing only odd winding numbers.
Moreover, if the system contains an odd number of particles and $\Omega$ includes a mixture of even and odd winding numbers, the resulting EC state does not have a definite parity under inversion. In such cases, the states have components in both parity subspaces. Consequently, the maximum fidelity achievable in a QOC protocol is limited by the overlap of the initial state with the two parity subspaces.

Our initial state $\ket{\Psi_0}$  has even parity under inversion symmetry (see Appendix \ref{ap:symmetricgs} for a detailed proof). Therefore, for those systems in which the restrictions due to the symmetry apply, only the symmetric (even parity) subspace is reachable via dipolar QOC. In this case, the upper bound for the fidelity is 
\begin{equation}
\label{eq:symbound}
    F_{\mathrm{max}} =  \sum_{\substack{k\in\Omega \\ kN \;\mathrm{even}} } \frac{1}{K} \, .
\end{equation}

\begin{figure*}[ht]
    \centering
    \includegraphics[width=0.99\linewidth]{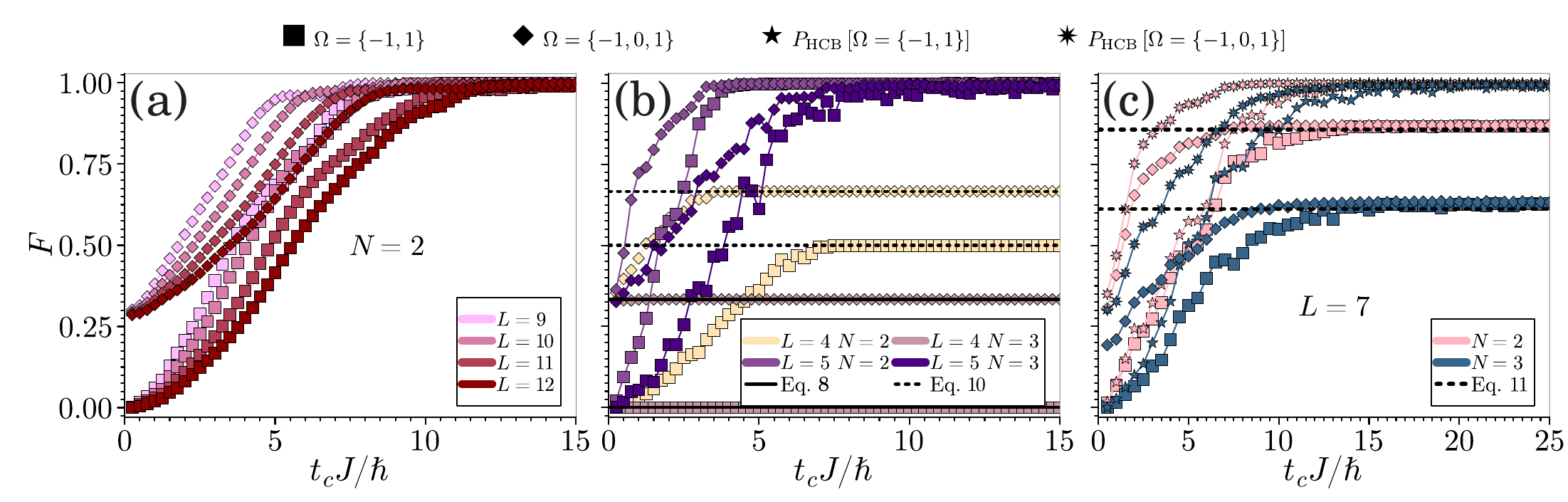}
    \caption{Final fidelity obtained via the dipolar QOC protocol as a function of the control time  $ t_c J /\hbar$ for the preparation of different entangled current states $\ket{\Psi_{\mathrm{EC}}}$. Each target state is represented with a different marker shape following the legend above the plots. Panel (a) shows systems with $N=2$ particles and a varying number of sites, while panel (b) presents systems with $N=2,3$ particles for only $L=4,5$ sites. These two panels consider $U/J = 1$ and $U_d / (d^3 J)=1.25$ interaction strengths. Results shown in panel (c) are for $L=7$ sites, with experimental $U/J=74$ and $U_d/(d^3 J)=1.11$ interaction strengths. The horizontal lines represent the maximum possible fidelity given by Eqs.~(\ref{eq:symbound}) and~(\ref{eq:boundDI}) in panel (b), and by Eq.~(\ref{eq:boundHCB}) in panel (c). Each point shows the best fidelity achieved over 10 optimization runs (to reduce the dependence on the initial trajectory proposed), with $M=30$ control pulses.}
    \label{fig:time}
\end{figure*}

An additional symmetry applies to systems with both even and odd numbers of sites: reflection across the $xy$ plane, corresponding to the transformation $z\to -z$. In contrast to inversion symmetry, this reflection does not affect the controllability of the system, as all states have even parity under this transformation.

In systems with an even number of sites and only two bosons, there exists an eigenstate of the dBHH, $\ket{\Psi_{\mathrm{DI}}}$, for any value of the orientation of $\boldsymbol{\mu}$. This specific state is of the form
\begin{equation}
    \ket{\Psi_{\mathrm{DI}}} = \frac{1}{\sqrt{2 L}} \sum_{j=1}^{L} (-1)^{j}  \,\creation_j \creation_j \ket{\operatorname{vac}} \, ,
    \label{eq:PsiDI}
\end{equation}
and its eigenvalue coincides with the on-site interaction strength $\lambda_{\mathrm{DI}} = U$. As an eigenstate of the Hamiltonian at any time, it constitutes a protected state whose projection onto the system's state will remain constant throughout the evolution. As a result, it is not possible to reach or modify this component via dipolar QOC, and the system's state will remain orthogonal to $\ket{\Psi_{\mathrm{DI}}}$ during the entire protocol. This reduces the controllability of certain systems. Notably, for $L=4, 8, 12, \dots$, the state $\ket{\Psi_{\mathrm{DI}}}$ is even under inversion symmetry, which implies that it lies within the symmetric subspace. Consequently, it prevents achieving a state inside the even subspace.
Then, the maximum fidelity that can be obtained when preparing an EC state in systems with $L=4, 8, 12, \dots$ and $N=2$ bosons is 
\begin{equation}
\label{eq:boundDI}
    F_\mathrm{max} = 1-\left|\frac{1}{\sqrt{K L}} \sum_{k\in \Omega} \left( \delta_{k, L/4} + \delta_{k, 3L/4} \right)\right|^2 \, .
\end{equation}

\section{\label{sec:Dipolarcontrol} Numerical implementation}

Having examined the limitations and the different upper bounds of the controllability, we have employed GRAPE algorithm implemented inside the \texttt{JuliaQuantumControl}~\cite{goerz_quantum_2022} framework to numerically find the optimal trajectories of the orientation of the dipole moment $\boldsymbol{\mu}(t)$. We extended the framework to allow us to use the dipolar control Hamiltonian [Eq.~(\ref{eq:dipolarcontrol})].
This framework is based on pulsed controls, where the control function $\boldsymbol{\theta}(t)$ is modeled as a piecewise-constant function defined on a discretized time grid of $M$ control steps of equal duration. We have chosen this numerical implementation as it is a highly efficient framework to simulate and optimize the evolution of controlled quantum systems, as demonstrated in related works \cite{goerz_quantum_2022, perciavalle_quantum_2024}. In the computations, we have optimized the dipolar trajectory using $M=30$ with an objective for a fidelity $F\geq 0.999$. We limited the number of timesteps to suppress high-frequency GRAPE oscillations while ensuring experimentally accessible times (see Sec. \ref{sec:experimental} for more details). 

Fig. \ref{fig:time} displays the best fidelity $F$ achieved over 10 optimization runs (see Appendix \ref{ap:optimizations} for convergence discussion) as a function of $t_c$ for various lattice sizes and boson numbers $N=2,3$. Panel (a) shows that all $N=2$ systems reach the threshold fidelity, demonstrating GRAPE's ability to locate optimal solutions where no state-preparation constraints exist.
In contrast, panel (b) shows that while the system with $L=5$ sites still achieves maximal fidelity, the $L=4$ system always saturates before reaching an accurate preparation of the state. The saturation values for $N=2$ and $N=3$ particles match those in Eq.~(\ref{eq:boundDI}) and Eq.~(\ref{eq:symbound}), respectively, due to dipolar-immune state blocking preparation in the former and inversion symmetry limiting evolution to even parity states in the latter. The saturation of all previously established upper bounds confirms that our analysis, based on linearly independent controls, accurately predicts the performance limits of dipolar QOC for the preparation of EC states. The absence of additional sources of controllability limitations in the numerical implementation reinforces the conclusion that symmetry constraints fully determine the controllability bounds in the preparation of EC states. As shown in Appendix \ref{ap:spatiallyentangled}, the preparation of different entangled states is also possible without controllability limitations.

Additionally, in all panels of Fig. \ref{fig:time}, each system exhibits a minimum control time where the maximum fidelity is reached, which increases with both the number of sites and bosons. From Fig. \ref{fig:time} (a), we observe that the scaling of the minimum control time is almost linear with $L$, making the system controllable in reasonable time frames as the system grows. This timescale arises due to the bounded nature of the control and also depends on the target state; greater initial overlap with the target reduces the required control time. In Appendix \ref{ap:qsl}, we analyze this feature and show that the protocol does not achieve the minimum control time predicted by the quantum speed limit.

\section{Hard Core bosons}\label{sec:hardcore}

To give an example using realistic interaction parameters, we have used the dBHH parameters from  \cite{baier_extended_2016}: $U/J= 74$ and $U_d/(d^3 J) = 1.11$. These values correspond to a regime of strong on-site interaction, which effectively sets the system in the dipolar hard-core bosons (HCB) regime, also observed in other studies \cite{su_dipolar_2023}. In this regime, states with an expected occupation greater than one per site are excluded, effectively reducing the Hilbert space to fermionic-like Fock states. This means that the family of EC states previously considered is no longer physically viable, as they include components outside the HCB subspace. In Fig. \ref{fig:time} (c), we have compared the preparation of the original $\ket{\Psi_{\mathrm{EC}}}$ states with the preparation of $P_{\mathrm{HCB}}\ket{\Psi_{\mathrm{EC}}}$ states, which are their projections onto the HCB subspace. From the figure, we can observe that the projected states can be reached with high fidelity even when the original states cannot be reached. Therefore, the performance of dipolar QOC with experimental-like parameters remains consistent with the results discussed above: every target state that is physically feasible and is not forbidden due to symmetry constraints can be prepared with high fidelity. In fact, the bound of the maximum achievable fidelity in the HCB limit for any EC current is given by: 
\begin{equation}
\label{eq:boundHCB}
    F_{\mathrm{max}}=\bra{\Psi_{\mathrm{EC}}} P_{\mathrm{HCB}} \ket{\Psi_{\mathrm{EC}}} = L^{-N} \frac{L!}{(L-N)!}\, .
\end{equation}
This limit does not precisely match our computations, as we have used a finite value for the on-site interaction $U/J$. Nevertheless, it provides a very close estimation of the behavior of the system. 

\section{Experimental feasibility}
\label{sec:experimental}

     The value of the hopping parameter in realizations of a dipolar Bose Hubbard system \cite{baier_extended_2016} is of the order of $J/h \sim 30$\,Hz, which, for the convergence times observed in numerical simulations, leads to control times of $t_c \sim 50$\,ms. In this case, each one of the $M=30$ timesteps used for the dipolar QOC protocol has a duration of $\tau_M\sim 1.5$\,ms. The rotation frequencies of the magnetic field achieved using coils are $\Omega \sim 2\pi\times 1$\,kHz \cite{tang_tuning_2018}, leading to a maximum time needed to update the magnetic field orientation of $\tau_{D}\sim 0.25$ ms (where only 90 degrees rotation is needed if it takes into account the inversion symmetry of the interaction). From the parameters chosen, the reorientation time is smaller than the duration of each of the steps, which means that the polarization schedule can be followed. This result leads to a maximum oscillation frequency $\omega_M = \pi/(2\tau_M) \sim 2\pi \times 167 $ Hz. As this is smaller than the maximum available frequency, it effectively suppresses the high-frequency oscillations common in GRAPE optimizations.
     
     From the previous estimations, we conclude that the specific step profile used as an example in the work [see Fig.~1(c)] can be approximated, although finite coil rise times prevent reproducing it exactly. For each specific experimental implementation, the polarization trajectory optimization can be modified to take into account the impedancy of the coils, geometry restrictions and the eddy currents induced in the metallic surroundings.

\section{\label{sec:Conclusion}Conclusion}

In this work, we have proposed the control on the orientation of the dipoles as a powerful tool to drive systems' evolution and perform high-fidelity state preparation for ultracold dipolar gases. We have identified the fundamental limits to fidelity imposed by symmetries, invariant states, and the hard-core boson regime, and verified numerically that these limits are attainable in ring lattices for the preparation of EC states. The absence of additional sources of controllability limitations beyond the symmetry constraints reinforces the conclusion that symmetry fully determines the controllability bounds, providing a practical blueprint.
Beyond the specific preparation of EC states, our results position dipolar optimal control as a plausible strategy for steering dipolar systems, further cementing their role in the development of quantum technologies.

Further work will examine the generalization of the dipolar QOC to other discrete systems, such as 2D lattice arrays and continuous rings. In those systems, the analysis of the symmetries will also play a key role in identifying the set of accessible states through this method.

\section*{Acknowledgments}

This project is funded by Grant PID2023-147475NB-I00 funded by  MICIU/AEI/10.13039/501100011033 and FEDER, UE,
by Grant No. 2021SGR01095
from Generalitat de Catalunya, and by grant CEX2024-001451-M funded by MICIU/AEI/10.13039/501100011033.
H. B.-M. is supported by FPI Grant PRE2022-104397 funded by MICIU/AEI/10.13039/501100011033 and by ESF+. 
F.I. acknowledges funding from ANID through FONDECYT Postdoctorado No. 3230023. 

\section*{Data Availability}
The data that support the findings of this article are openly available \cite{briongos_merino_data_2025}.

\bibliography{dipolarcontrol}

\appendix

\begin{figure*}[tb]
    \centering
    \includegraphics[width=0.95\linewidth]{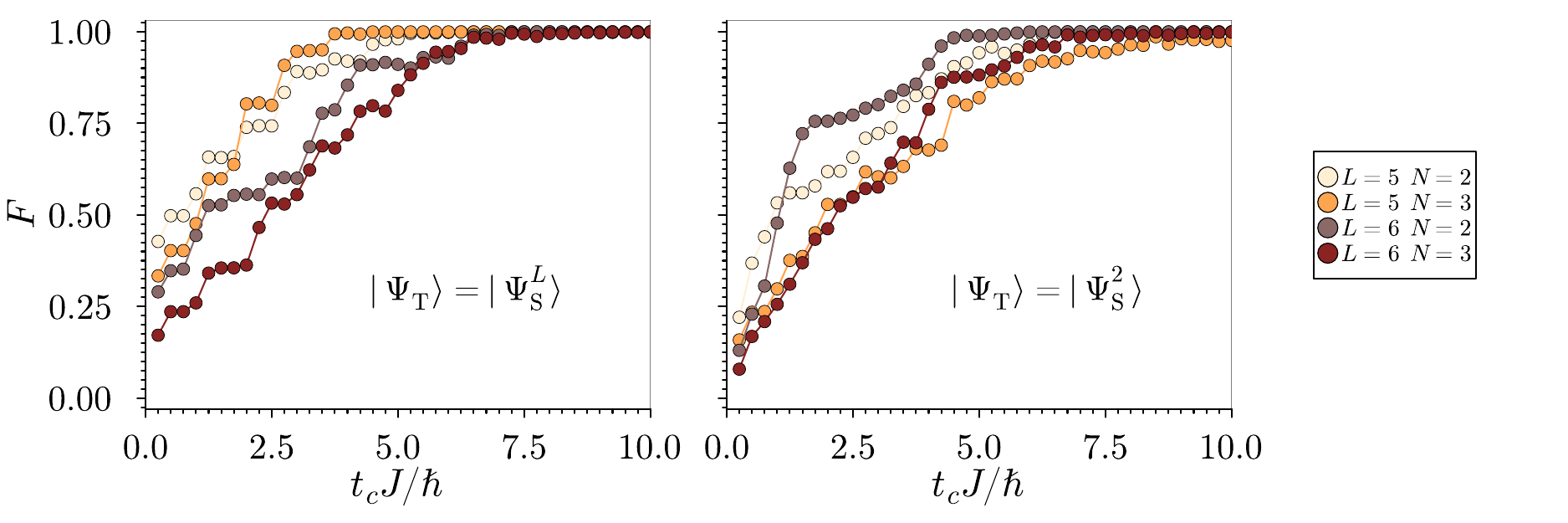}
    \caption{Final fidelity obtained with the dipolar QOC protocol as a function of the control time  $ t_c J /\hbar$ for the target states indicated in each panel. Each color corresponds to distinct combinations of $N$ and $L$. Each point shows the best fidelity achieved over 10 optimization runs using $M=30$ control steps. The results shown in both panels correspond to interaction strengths of $U/J = 1$ and $U_d / (d^3 J)=1.25$.}
    \label{fig:SupMat}
\end{figure*}

\section{Production of spatially entangled states}
\label{ap:spatiallyentangled}

Besides the EC states, we have also performed calculations for other entangled targets. We report here the numerical production of spatially entangled states as complementary evidence of the possibilities of the protocol. Examples of spatially entangled states for a system with $L$ sites and $N$ bosons are 
\begin{equation}
\label{eq:supmat}
    \ket{\Psi_{\mathrm{S}}^L} = \frac{1}{\sqrt{L}}\sum_{j=1}^L \; \prod_{l=0}^{N-1} \creation_{j+l} \; \ket{\mathrm{vac}} \;,
\end{equation}
and 
\begin{equation}
\label{eq:supmat2}
    \ket{\Psi_{\mathrm{S}}^2} = \frac{1}{\sqrt{2}}\left[ \; \prod_{j=1}^{N} \creation_{j} + \prod_{j=1}^{N} \creation_{\lceil L/2 \rceil + j}\right] \; \ket{\mathrm{vac}} \; ;
\end{equation}
which are superpositions of Fock states where the bosons are placed in $N$ consecutive sites. Both states are even under inversion symmetry whenever it is present in the system, therefore both of them should be reachable with the proposed driving. Moreover, because no site hosts more than a single particle, the overlap with the DI [eq. \ref{eq:PsiDI}] state vanishes, and both are achievable in the HCB regime.
Fig. \ref{fig:SupMat} shows the fidelity achieved using the dipolar QOC protocol to prepare the spatially entangled states (\ref{eq:supmat}) and (\ref{eq:supmat2}). Similar to the preparation of the EC states, the protocol requires a minimum control time to achieve maximum fidelity.

\section{Ground-state parity}
\label{ap:symmetricgs}

To prove that the ground state of the dBHH in a ring lattice with an even number of sites is even under inversion symmetry, we start by using the Perron-Frobenius theorem for positive symmetric matrices \cite{ninio_simple_1976}. This theorem states that if $\hat{A}=(a_{ij})$ is an $n\times n$ real symmetric matrix with all elements $a_{ij}\geq0$, then its largest eigenvalue $\lambda$ is positive and non-degenerate and has an associated eigenvector with positive components. For a given dipolar Bose-Hubbard Hamiltonian $\hat{H}_{\mathrm{dBH}}(\boldsymbol{\mu})$ we construct such a matrix by defining $\hat{A} = C\mathbb{I} - \hat{H}_{\mathrm{dBH}}(\boldsymbol{\mu})$ written in the Fock basis $\{\ket{n_1,n_2,\dots,n_{L}}\}$, with $C$ a positive constant larger than any diagonal element of $\hat{H}_{\mathrm{dBH}}(\boldsymbol{\mu})$ for any $\boldsymbol{\mu}$.

As all diagonal elements are positive by construction and all off-diagonal elements associated with nearest-neighbor tunneling are also positive due to the minus sign in the construction of $\hat{A}$ (which effectively inverts the sign of the hopping terms), this operator in matrix representation is real, positive and symmetric. Then, the theorem implies that the largest eigenvalue $\lambda$ of $\hat{A}$ is positive and non-degenerate. Moreover, the eigenvector corresponding to the largest eigenvalue of $\hat{A}$ is precisely the eigenvector corresponding to the smallest (ground-state) eigenvalue of $\hat{H}_{\mathrm{dBH}}(\boldsymbol{\mu})$, and this eigenvector has all positive entries.

The ground state is written in the Fock basis as $\ket{\Psi_{\mathrm{GS}}}=\sum c_{n_1, n_2, \dots, n_{L}} \ket{n_1, n_2, \dots, n_{L}}$, where all coefficients $c_{n_1, n_2, \dots, n_{L}}$ are real and positive. The inversion symmetry changes the labeling of the sites as $j \to j+L/2$ (mod $L$), and therefore the eigenstate transforms as $\hat{I}\ket{\Psi_{\mathrm{GS}}}=\sum  c_{n_{\pi(1)}, n_{\pi(2)}, \dots, n_{\pi(L)}} \ket{n_1, n_2, \dots, n_{L}}$, where $\pi$ is the permutation that maps site $j$ to $j+L/2 \pmod{L}$. This state is an eigenvector of the symmetry because $[\hat{I}, \hat{H}_{\mathrm{dBH}}(\boldsymbol{\mu})]=0$. Since $\ket{\Psi_{\mathrm{GS}}}$ is the unique ground state and all its coefficients are positive, and knowing that $\hat{I}$ is a unitary operator such that $\hat{I}\hat{I}=\mathbb{I}$, its eigenvalue $\lambda_I$ must be $\pm 1$. To satisfy the eigenvalue equation $\hat{I}\ket{\Psi_{\mathrm{GS}}} = \lambda_I \ket{\Psi_{\mathrm{GS}}}$, implies that $c_{n_{\pi(1)}, \dots, n_{\pi(L)}} = \lambda_I \, c_{n_1, \dots, n_{L}}$. Since all coefficients are positive, $\lambda_I$ cannot be $-1$ and thus $\lambda_I=1$. Therefore, the ground state of the dBHH is even under inversion symmetry in any ring system that presents spatial inversion invariance.

\begin{figure}[tb]
    \centering
    \includegraphics[width=0.99\linewidth]{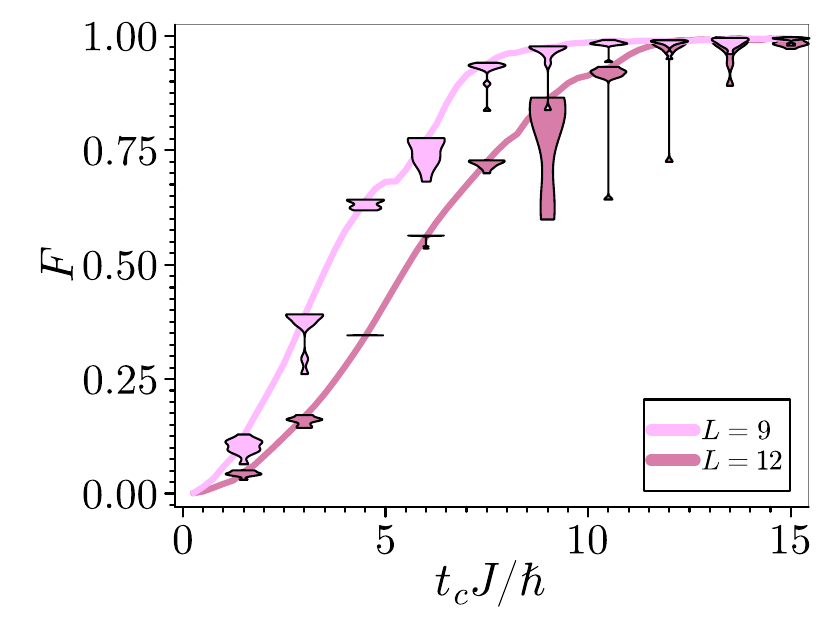}
    \caption{Final fidelity as a function of the control time  $ t_c J /\hbar$ for preparing the EC state with $\Omega=\{-1,1\}$.  The violin plots represent the dispersion of the optimization results over 10 optimization runs using $M=30$ control steps. The results shown correspond to optimizations from Fig. \ref{fig:time} (a).}
    \label{fig:SupMatconvergence}
\end{figure}

\section{Optimization trials and convergence}
\label{ap:optimizations}

The Gradient-Ascent Pulse Engineering (GRAPE) algorithm is a highly effective tool for quantum optimal control  \cite{goerz_quantum_2022, perciavalle_quantum_2024}. As it relies on local, gradient-based updates, it is exposed to convergence issues, most notably the risk of becoming trapped in suboptimal local maxima within complex, many-body control landscapes. This characteristic trapping is evident in the optimization data presented in Fig. \ref{fig:SupMatconvergence}, where a subset of individual optimization runs prematurely converges to lower fidelity values, resulting in elongated lower tails in the violin plots. 

However, as illustrated by the dense clustering of the individual realizations immediately beneath the maximum-fidelity envelope, the optimization landscape is dominated by favorable regions that converge to the global optimum. As the probability density distributions show, the parameter space basin leading to the maximum fidelity is large. Consequently, this strong clustering demonstrates that executing only a small ensemble of optimizations with random initial seeds is sufficient to find an optimal trajectory that reaches the maximum available fidelity.

\section{Quantum speed limit}
\label{ap:qsl}

\begin{figure}[bt]
    \centering
    \includegraphics[width=0.99\linewidth]{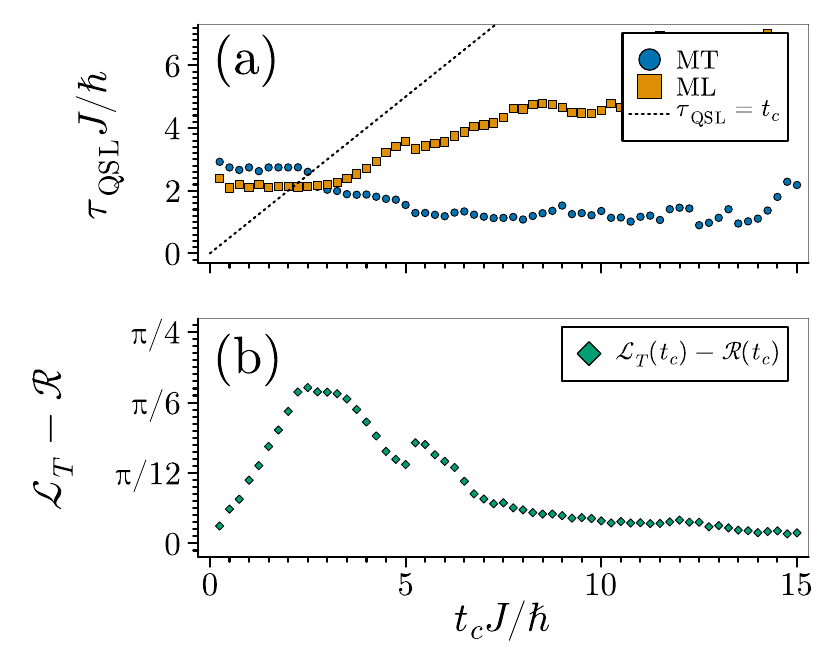}
    \caption{Evaluation of  (a) the quantum speed limit and (b) the triangle inequality for the optimal trajectories obtained for the system with $L=9$, $N=2$ for the preparation of the EC state with $\Omega=\{-1,1\}$. }
    \label{fig:qsl}
\end{figure}

The quantum speed limit bounds for time-dependent Hamiltonians \cite{deffner_energy_2013, okuyama_comment_2018}, as it is in the case of our dipolar control protocol, read 
\begin{equation}
\label{eq:qslclassic}
\tau_{\text{QSL}} \geqslant \max \left\{ \frac{\hbar \mathcal{L}(\Psi_0, \Psi_{\text{T}})}{\Delta H_\tau} , \frac{4\hbar \mathcal{L}^2 (\Psi_0, \Psi_{\text{T}}) }{\pi^2 \tilde{E}_\tau} \right\},
\end{equation}
where $\mathcal{L}(\Psi_0, \Psi_{\text{T}}) = \operatorname{arccos}(|\langle \Psi_0 | \Psi_{\text{T}}\rangle|)$ is the Fubini-Study distance between initial and target states, $\Delta H_\tau = \frac{1}{\tau}\int_{0}^{\tau} dt\,\sqrt{\langle \hat{ \mathcal{H}}^2(t)\rangle - \langle \hat{\mathcal{H}}(t) \rangle^2 }$ is the time-averaged variance of the Hamiltonian and $\tilde{E}_\tau = \frac{1}{\tau} \int_0^\tau dt \, |\langle \Psi_0 | \hat {\mathcal{H}'}(t)|\Psi(t)\rangle|$ is the time average of the transition matrix elements of the regularized Hamiltonian. The two terms are known as Mandelstam-Tamm (MT) and Margolus-Levitin (ML) bounds, respectively. The regularized Hamiltonian $\hat{\mathcal{ H}}'(t)$ is the time-dependent Hamiltonian with a zero-energy ground state at each time. 

To complement the previous bounds, one can compute the residual distance difference $\mathcal{R}(t_c)$
and compare it with the distance to target $\mathcal{L}_T(t)=\mathcal{L}(\Psi(t), \Psi_{\text{T}})$. In the cases where the evolution follows a \textit{quantum brachistochrone} (the time-optimal geodesic path)~\cite{carlini_time-optimal_2006, jones_geometric_2010}, both quantities match. Otherwise, they fulfill the following triangle inequality:
\begin{equation}
\label{eq:qslgeodesic}
\mathcal{R} (t_c) = \mathcal{L} (\Psi_0, \Psi_{\text{T}}) -  \mathcal{L} (\Psi_0, \Psi(t_c)) \leq \mathcal{L}_T(t_c)\; .
\end{equation}

As a representative example, Fig. \ref{fig:qsl} shows the value of the quantum speed limits [Eq. \ref{eq:qslclassic})] and the triangle inequality [Eq. (\ref{eq:qslgeodesic})] for the optimal trajectories that achieve the fidelity shown in Fig. \ref{fig:time} (a) for the system with $L=9$, $N=2$ in the preparation of the EC state with $\Omega=\{-1,1\}$. In panel (a), it is shown that the MT and ML bounds give $\tau_{\mathrm{QSL}}$ a value no higher than $6\;J/\hbar$. In fact, they cross the line $\tau_{\mathrm{QSL}}=t_c$ at a value of approximately $2\;J/\hbar$. The latter implies that, for the amount of energy pumped into the system, the state preparation task could ideally be completed for all $t_c \geq 2\;J/\hbar$. Nevertheless, from Fig. \ref{fig:time} (a), one can observe that the control time needed with the dipolar QOC is of the order of $10 \;J/\hbar$, indicating that the protocol does not achieve the task in the minimum possible time. Moreover, panel (b) shows that the triangular inequality is only nearly saturated for control times larger than $t_c=10\,J/ \hbar$.  These correspond to control protocols long enough to drive the system to the target state, which by construction lies along the geodesic path. Therefore, the dipolar QOC protocol does not drive the system across the optimal path in the Hilbert space and requires more control time than the value predicted by Eq. (\ref{eq:qslclassic}) to complete the state preparation task.

\section{Fidelity bounds}

We include here the explicit derivation of the fidelity bounds presented in the main text. To obtain the bounds we mainly use the relation between the site ($\anihilation$, $\creation$) and momentum ($\banihilation$, $\bcreation$) operators, and the action of the inversion operator $\hat I$ over the EC states. 

    To compute the symmetry bound, we create the projector into the symmetric subspace $\hat P_S = \frac{1+\hat I}{2} $ via the inversion operator. Then, the maximum achievable fidelity is computed as the expected value of the projector with the given target state.
    \begin{align*}
        F_{\mathrm{max}}&=\bra{\Psi_{\mathrm{EC}}}\hat P_S \ket{\Psi_{\mathrm{EC}}} \\ =& \frac{1}{2 K N !} \sum_{\substack{k\in\Omega \\ k'\in\Omega}} \left(1 + e^{-i\pi N k} \right) \bra{\operatorname{vac}} \left(\hat{b}_{k'}\right)^{N}  \left(\hat{b}_{k}^{\dagger}\right)^{N} \ket{\operatorname{vac}} \\ 
        =&\frac{1}{2 K N !} \sum_{\substack{k\in\Omega \\ k'\in\Omega}} \left(1 + e^{-i\pi N k} \right) \delta_{k,k'} N! = \sum_{\substack{k \in \Omega \\ kN \; \mathrm{even}}} \frac{1}{K}
    \end{align*}
    
        The state of the system will never have overlap with $\ket{\Psi_{\mathrm{DI}}}$ when starting the QOC protocol from the ground state of the system. Then, to compute the bound due to the DI eigenstate, it is enough to subtract the overlap of the DI eigenstate with the target EC state, as there is no symmetry limitation to the production of EC states in systems with two particles. Writting the DI eigenstate in terms of momentum operators the overlap reads

\begin{widetext}
    \begin{align*}
        \bra{\Psi_{\mathrm{DI}}}\ket{\Psi_{\mathrm{EC}}} &= \frac{1}{\sqrt{4K L}} \bra{\operatorname{vac}} \sum_{j=1}^{L} (-1)^{j}  \frac{1}{L} \sum_{p=0}^{L-1} \sum_{p'=0}^{L-1} e^{i2\pi (p + p') j / L} \banihilation_p \banihilation_{p'}  \sum_{k\in\Omega} \left(\hat{b}_k^\dagger\right)^{2} \ket{\operatorname{vac}} \\
        &= \frac{1}{\sqrt{4K L^3}} \sum_{j=1}^{L} (-1)^{j}  \sum_{p=0}^{L-1} \sum_{p'=0}^{L-1} e^{i2\pi (p + p') j / L} \sum_{k\in\Omega} \bra{\operatorname{vac}} \banihilation_p \banihilation_{p'} \left(\hat{b}_k^\dagger\right)^{2} \ket{\operatorname{vac}} = \frac{1}{\sqrt{K L^3}} \sum_{k\in\Omega} \sum_{j=1}^{L} (-1)^{j} e^{i4\pi k j / L} \\ &= \frac{1}{\sqrt{K L^3}} \sum_{k\in\Omega} \sum_{j=1}^{L} e^{i2\pi (2k - L/2) j / L} = \frac{1}{\sqrt{K L}} \sum_{k\in\Omega} (\delta_{k,L/4}+\delta_{k,3L/4})  \; .
    \end{align*}
\end{widetext}
    Thus, the maximum attainable fidelity with the dipolar control is
    \begin{align*}
        F_\mathrm{max} &= 1-|\bra{\Psi_{\mathrm{DI}}} \Psi_{\mathrm{EC}}\rangle|^2 \\ &= 1-\left|\frac{1}{\sqrt{K L}} \sum_{k\in \Omega} \left( \delta_{k, L/4} + \delta_{k, 3L/4} \right)\right|^2 \, .
    \end{align*}

    To compute the bound due to the HCB regime, we first write the EC states in terms of the $\anihilation$ and $\creation$ operators    
\begin{widetext}
    \begin{equation*}
    \begin{aligned}
        \ket{\Psi_{\mathrm{EC}}} &= \frac{1}{\sqrt{K N !}}\sum_{k\in\Omega} \left(\hat{b}_k^\dagger\right)^{N} \ket{\operatorname{vac}} =\frac{L^{-N/2}}{\sqrt{K N !}}\sum_{k\in\Omega} \left( \sum_{j=1}^{L} e^{i 2\pi k j / L} \hat{a}_j^\dagger \right)^{N} \ket{\operatorname{vac}} \\
        &=\frac{L^{-N/2}}{\sqrt{K N !}}\sum_{k\in\Omega} \left[ \sum_{\substack{l_1,l_2,\dots,l_{L}\\l_1+l_2+\dots+l_{L}=N}} 
        \begin{pmatrix}
            N \\
            l_1,l_2, \ldots, l_{L}\end{pmatrix}
            \prod_{r=1}^{L} \left( e^{i\frac{2\pi k r}{L}} \creation_{r}\right)^{l_r} \right] \ket{\operatorname{vac}} \; ;\\
        P_{\mathrm{HCB}}\ket{\Psi_{\mathrm{EC}}} 
        &=\frac{L^{-N/2}}{\sqrt{K / N !}}\sum_{k\in\Omega}\left[ \sum_{\substack{l_1,l_2,\dots,l_{L} = \{0, 1\}\\l_1+l_2+\dots+l_{L}=N}}
            \prod_{r=1}^{L} \left( e^{i\frac{2\pi k r}{L}} \creation_{r}\right)^{l_r} \right] \ket{\operatorname{vac}}  .
    \end{aligned}
    \end{equation*}

    Where the last line is the projection into HCB regime, that only has Fock states with one boson maximum per well. Therefore, only allowed values for the $l_r$ indices are 0 and 1.
    Finally, the maximum reachable fidelity will be the norm of the projected state
    
    \begin{equation*}
    \begin{aligned}
        &F_{\mathrm{max}} = \bra{\Psi_{\mathrm{EC}}}P_{\mathrm{HCB}}^\dagger P_{\mathrm{HCB}}\ket{\Psi_{\mathrm{EC}}}= \\
        &= \frac{L^{-N} N!}{K} \bra{\operatorname{vac}} \sum_{k', k \in\Omega} \left[ \sum_{\substack{l'_1,l'_2,\dots,l'_{L} = \{0, 1\}\\l'_1+l'_2+\dots+l'_{L}=N}}
        \prod_{r'=1}^{L} 
        \left( e^{-i\frac{2\pi k' r'}{L}} \anihilation_{r'}\right)^{l'_{r'}} \right] \left[
        \sum_{\substack{l_1,l_2,\dots,l_{L} = \{0, 1\}\\l_1+l_2+\dots+l_{L}=N}} 
        \prod_{r=1}^{L} \left( e^{i\frac{2\pi k r}{L}} \creation_{r}\right)^{l_r} \right] \ket{\operatorname{vac}} 
         \\
        &=  \frac{L^{-N} N!}{K}  \sum_{k', k \in\Omega} \sum_{\substack{l'_1,l'_2,\dots,l'_{L} = \{0, 1\}\\l'_1+l'_2+\dots+l'_{L}=N}}
        \sum_{\substack{l_1,l_2,\dots,l_{L} = \{0, 1\}\\l_1+l_2+\dots+l_{L}=N}}
        \bra{\operatorname{vac}}
        \prod_{r'=1}^{L} 
        \prod_{r=1}^{L}
        e^{-i\frac{2\pi (k' r' l'_{r'} - k r l_r)}{L}}  \anihilation_{r'}^{l'_{r'}} 
        \hat{a}_{r}^{\dagger l_r}\ket{\operatorname{vac}} 
         \\
        &=  \frac{L^{-N} N!}{K}  \sum_{k', k \in\Omega} \sum_{\substack{l'_1,l'_2,\dots,l'_{L} = \{0, 1\}\\l'_1+l'_2+\dots+l'_{L}=N}}
        \sum_{\substack{l_1,l_2,\dots,l_{L} = \{0, 1\}\\l_1+l_2+\dots+l_{L}=N}}
        \prod_{r'=1}^{L} 
        \prod_{r=1}^{L}
        e^{-i\frac{2\pi (k' r' l'_{r'} - k r l_r)}{L}} \delta_{r,r'} \delta_{l_r,l'_r}
         \\
        &=  \frac{L^{-N} N!}{K}  \sum_{k', k \in\Omega} 
        \sum_{\substack{l_1,l_2,\dots,l_{L} = \{0, 1\}\\l_1+l_2+\dots+l_{L}=N}} 
        \prod_{r=1}^{L}
        e^{-i\frac{2\pi (k' - k) r l_r}{L}} = \frac{L^{-N} N!}{K} \sum_{k', k \in\Omega}  \frac{\delta_{k,k'}  L!}{N!(L-N)!} = L^{-N} \frac{L!}{(L-N)!} \; ,
    \end{aligned}
    \end{equation*}

    where we have used that     
    \begin{equation*}
        \sum_{\substack{l_1,l_2,\dots,l_{L} = \{0, 1\}\\l_1+l_2+\dots+l_{L}=N}} 
        1 = \frac{L!}{N!(L-N)!} \; .
    \end{equation*}
        
\end{widetext}

\end{document}